\documentclass[journal]{IEEEtran}
\usepackage{epsfig}
\usepackage{graphicx}
\usepackage{amsmath}
\usepackage{amssymb}
\usepackage{relsize}

\usepackage{verbatim}
\usepackage{hyperref}

\usepackage{color}
\usepackage[table]{xcolor}

\usepackage{cite}
\newcommand{\etal}{\textit{et al.}}
\usepackage{graphicx}
\usepackage{siunitx}
\usepackage{booktabs}
\usepackage{bm}

\definecolor{blue}{rgb}{0,0,0}  
\definecolor{bluedec}{rgb}{0,0,0}  

\usepackage{cite}

\ifCLASSINFOpdf
\else
\fi
\begin{document}
%
\title{A Coarse-to-fine Deformable Transformation Framework For Unsupervised Multi-contrast MR Image Registration With Dual Consistency Constraint}
%
%
%

\author{Weijian Huang, Hao Yang, Xinfeng Liu, Cheng Li, Ian Zhang, Rongpin Wang, Hairong Zheng, \IEEEmembership{Senior Member, IEEE}, and Shanshan Wang, \IEEEmembership{Senior Member, IEEE}

\thanks{Manuscript received October 1, 2020; accepted February 09, 2021. This research was partly supported by Scientific and Technical Innovation 2030-"New Generation Artificial Intelligence" Project (2020AAA0104100, 2020AAA0104105), the National Natural Science Foundation of China (61871371, 81830056), Key-Area Research and Development Program of GuangDong Province (2018B010109009), the Basic Research Program of Shenzhen (JCYJ20180507182400762), Youth Innovation Promotion Association Program of Chinese Academy of Sciences (2019351).(Corresponding author: S. Wang)}
\thanks{W. Huang, H. Yang, L. Cheng, I. Zhang, H. Zheng, and S. Wang are with Paul C. Lauterbur Research Center for Biomedical Imaging, Shenzhen Institutes of Advanced Technology, Chinese Academy of Sciences, China. S. Wang is also with Pengcheng Laboratory, Shenzhen, Guangdong, China and Pazhou Lab, Guangzhou, China (email: wj.huang@siat.ac.cn; hao.yang@siat.ac.cn; cheng.li6@siat.ac.cn; ianzhangfpv@gmail.com; hr.zheng@siat.ac.cn; sophiasswang@hotmail.com).}
\thanks{X. Liu and R. Wang are with Guizhou Provincial People's Hospital, Radiology Guiyang, Guizhou, China(email: lainiu6715613@163.com; wangrongpin@126.com).}}

\maketitle

\begin{abstract}
Multi-contrast magnetic resonance (MR) image registration is useful in the clinic to achieve fast and accurate imaging-based disease diagnosis and treatment planning. Nevertheless, the efficiency and performance of the existing registration algorithms can still be improved. In this paper, we propose a novel unsupervised learning-based framework to achieve accurate and efficient multi-contrast MR image registrations. Specifically, an end-to-end coarse-to-fine network architecture consisting of affine and deformable transformations is designed to improve the robustness and achieve end-to-end registration. Furthermore, a dual consistency constraint and a new prior knowledge-based loss function are developed to enhance the registration performances. The proposed method has been evaluated on a clinical dataset containing 555 cases, and encouraging performances have been achieved. Compared to the commonly utilized registration methods, including VoxelMorph, SyN, and LT-Net, the proposed method achieves better registration performance with a Dice score of 0.8397$\pm$0.0756 in identifying stroke lesions. With regards to the registration speed, our method is about 10 times faster than the most competitive method of SyN (Affine) when testing on a CPU. Moreover, we prove that our method can still perform well on more challenging tasks with lacking scanning information data, showing the high robustness for the clinical application.
\end{abstract}


%
\IEEEpeerreviewmaketitle

\vspace{-0.2cm}

\section{Introduction}
\label{sec:introduction}
Multi-modal medical imaging plays an important role in many clinical applications \cite{krebs2019learning, balakrishnan2018unsupervised, dalca2018unsupervised, balakrishnan2019voxelmorph, dalca2019unsupervised,Wang2017Learning, chaisaowong2018automated, marstal2019continuous, cao2018region,cao2018deformable,zhao2017novel,konig2015parallel,li2017pixel}. Among them, multi-contrast magnetic resonance (MR) imaging is one of the most prevalent techniques as different MR imaging sequences can provide versatile information and highlight different regions of interest of the patient. For instance, diffusion-weighted imaging (DWI) and apparent diffusion coefficient (ADC) are functional MR images based on the movement of water molecules \cite{stejskal1965spin, basser1994mr}. T1-weighted images (T1), T2-weighted images (T2), and fluid-attenuated inversion-recovery (FLAIR) are structural MR images \cite{lauterbur1973image} which can indicate different characteristics of anatomical structures. 

Multi-contrast MR imaging is of great significance for disease diagnosis and treatment response monitoring in clinical practices\cite{hill2001medical, brown1992survey, van1999automated, dawant2002non}. Structural MR images can clearly show the structures and boundaries of brain tissues but have moderate performances on discriminating brain lesions. On the other hand, functional MR images possess excellent capabilities of highlighting brain diseases, such as ischemic lesion regions. Analysis with multi-contrast images contributes to the comprehensive understanding of the patient. However, misalignment exists between different contrast images due to various issues of the scanning process, including physiological activities and eddy currents \cite{reese2003reduction}. Fig. 1 shows the multi-contrast images of four examples. Physical space alignment has been conducted utilizing the provided scanning information. However, misalignment between the different contrast images can still be observed. In some cases, the scanning information might be lost due to data storage or transfer, large misalignment can happen. Misalignment brings difficulties to identify lesions accurately, which may have adverse effects on disease diagnosis. Multi-contrast MR image registration is needed.

\begin{figure}[htbp]
\centering
\includegraphics[width=1\linewidth]{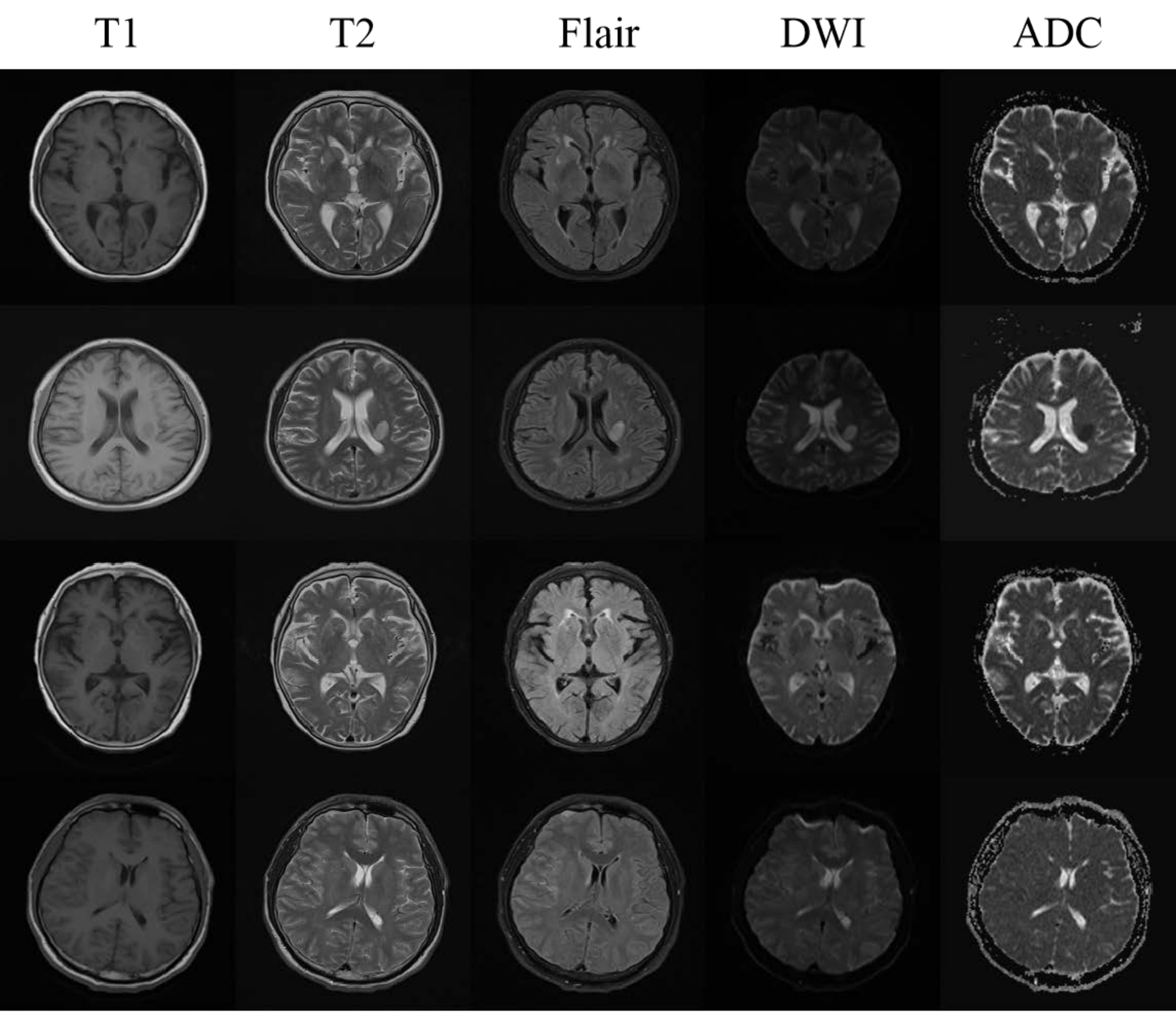}
\caption{MR multi-contrast brain images acquired from four candidates. There are differences between the contrasts that need to be registered.}
\label{Fig.1}
\end{figure}

Registration methods are available to alleviate the mismatch problem. Traditional multi-contrast registration algorithms rely on the interactive optimization process, which is not very applicable to the time-sensitive diagnosis required in clinical practices. Deep learning-based methods have been developed recently and can speed up the registration process at the cost of registration accuracy. 

To achieve accurate and fast multi-contrast MR image registration, this paper proposes a novel concise registration framework. Specifically, we have made the following contributions:

\begin{enumerate}
\item[1)] We propose an unsupervised coarse-to-fine registration framework. A coarse registration is obtained by an affine transformation network, which is then refined by a subsequent deformable transformation network. These two transformations are integrated, and end-to-end image registration is achieved. 

\item[2)] A dual consistency constraint is designed to maximize the cross-correlation of topology maps of multi-contrast MR images. The inverse deformation field is generated from the forward deformation field directly to reduce the time requirement. The designed consistency constraint is enforced on the bi-directional deformations so as to suppress pixel folding. 

\item[3)] A prior knowledge-based loss function is designed to improve the sensitivity of mutual information (MI) for more accurate registration. Specifically, a negative area constraint is designed to limit signals that are registered in the fixed images background.

\item[4)] Extensive experiments with or without the first step of physical space alignment show the superiority of the proposed registration method compared to existing widely-employed approaches.
\end{enumerate}

The rest of this paper is organized as follows: Section \ref{sec:related work} introduces related work in medical image registration, Section \ref{sec:method} describes our methods, Section \ref{sec:experiments and results} presents the experimental results and relevant analysis, and Section \ref{sec: conclusion} gives the conclusion.

\vspace{-0.25cm}
\section{Related Work}
\label{sec:related work}

\subsection{Conventional image registration methods}
Traditional image registration algorithms, such as elastic \cite{bajcsy1989multiresolution, shen2002hammer}, fluid \cite{beg2005computing, hart2009optimal, vercauteren2009diffeomorphic, chen2013large, wulff2015efficient} or B-spline models \cite{rueckert1999nonrigid}, are usually based on the iterative numerical solution of the optimization problem.Especially, in 1998, Thirion \etal \cite{thirion1998image} proposed a method called demons to estimate the velocity vector field between two adjacent images in a video. Specifically, they calculated the optical flow, used Gaussian filter to smooth the flow map, and optimized the predictions on each pair of images through multiple iterations. Since the successful implementation of demons, many variants were developed, such as the works by Wang \etal and Vercauteren \etal \cite{wang2005validation, vercauteren2007non}. In 2005, Beg \etal \cite{beg2005computing} proposed another famous registration algorithm, LDDMM (Large Displacement Diffeo-morphic Metric Mapping), by deducing and implementing the Euler-Lagrangian optimization to compute particle flows, solving a global variational problem, and estimating metrics for images. Subsequently, variants of this algorithm were also proposed, including Region-specific Diffeomorphic Metric Mapping (RDMM), vector momentum-parameterized Stationary Velocity Field (vSVF), and Symmetric image Normalization (SyN) \cite{shen2019region,shen2019networks,avants2008symmetric}. Among them, SyN \cite{avants2008symmetric} has been the most widely employed algorithm in medical image registration.It described an Euler-Lagrange optimization based symmetric image normalization method for maximizing the cross-correlation. Nevertheless, the efficiency of these methods can still be improved since these methods are based on iterative optimization \cite{balakrishnan2019voxelmorph,krebs2018learning}.

\subsection{Deep learning-based unimodal image registration}
With the fast development in the deep learning field, some deep learning-based image registration models have been proposed. Initially, deep learning was employed to enhance the registration performance of the iterative methods. Then, deep reinforcement learning was introduced to predict steps of transformations until the optimal alignment was reached \cite{krebs2017robust,liao2017artificial,ma2017multimodal,miao2018dilated}.
With the increased demand on the registration speed, deep learning-based registration methods were proposed \cite{de2019deep,li2018non,balakrishnan2018unsupervised,haskins2020deep}. One representative work in this group is STN (Spatial Transform Network), which generates dense deformable transformations to register images. Since then, STN has been modified and utilized in various situations \cite{jaderberg2015spatial}. Yoo \etal \cite{yoo2017ssemnet} successfully employed STN to register electron microscopy images. They trained an autoencoder to reconstruct the fixed images and calculated a new loss between the reconstructed fixed images and the corresponding moving images. Krebs \etal \cite{krebs2018learning,krebs2018unsupervised} proposed a random latent space learning method to alleviate the requirement on spatial regularization. De Vos \etal \cite{de2019deep} developed a multi-stage and multi-scale approach to register unimodal images with a normalized cross correlation (NCC) loss and a bending energy regularization. However, this approach cascaded multiple networks, which severely increased the computational complexity. Balakrishnan \etal proposed the famous framework, VoxelMorph, and its derivative versions \cite{balakrishnan2019voxelmorph, balakrishnan2018unsupervised, dalca2018unsupervised, dalca2019unsupervised}, which computed gradients of the transformation to backpropagate deformation errors during optimization. However, since the above methods all focus on unimodal image registration, multi-contrast image registration remains to be explored.

\subsection{Deep learning-based multi-modal image registration}
Since multi-contrast MR image registration is similar to multi-modal medical image registration, we discuss multi-modal registration in this section to give a more comprehensive description. Compared with unimodal registration, multi-modal registration is more challenging because it is difficult to define effective similarity measures to guide local matching across different modalities. Mutual information (MI) is the most frequently utilized supervision in existing studies\cite{maes1997multimodality}. Li \etal \cite{li2018multi} registered multi-modal retinal images by using the descriptor matching on the average phase map for global registration and using a deformable modality independent neighborhood descriptor method to locally optimize the registration results. Unfortunately, this method was based on manually designed features and it has limited robustness. Ceranka \etal \cite{ceranka2018registration} proposed a whole-body DWI and T1-weighted image registration method. This method roughly aligned the pelvis regions of the two modal images and then used MI to guide global registration. Cao \etal \cite{cao2017dual} developed an image synthesis-based method. They adopted a random forest to learn the transformation between computed tomography (CT) images and MR images, and synthesized pseudo CT images and pseudo MR images with similar anatomical structures. In this way, they transferred the multi-modal image registration task to a unimodal image registration task. Improved models over this original implementation were also proposed in \cite{cao2018deep}. Nonetheless, these methods require a robust domain transformation algorithm and their registration performances can be highly affected by the quality of the synthesized images \cite{cao2017dual}.
\section{Method}
\label{sec:method}
In this paper, we propose a concise registration algorithm for unsupervised multi-contrast MR image registration. The proposed method embeds an affine transformation network in a deformable network to achieve coarse-to-fine registrations. A dual consistency constraint is designed to further enhance the registration performance. Meanwhile, a prior knowledge-based guidance function is implemented. Here, let $K \in R$ represents the sample count in the multi-contrast datasets and $F \supset\left\{f^{1}, f^{2} \cdots f^{K}\right\}$ and $M \supset\left\{m^{1}, m^{2} \cdots m^{K}\right\}$ refer to the paired fixed image sets and moving image sets.

\subsection{Affine transformation network – ATNet}
STN \cite{jaderberg2015spatial} is a dynamic mechanism that can transform images or feature maps in a voxel-based manner. With this mechanism, a specific transformation can be performed all over the entire feature map, including scaling, cropping, rotating, etc. Owing to its high effectiveness, STN has been widely applied to deep learning-based registration tasks. 

We use STN to perform affine transformation on the moving images \cite{dong2002affine}, which geometrically consists of a non-singular linear transformation (transformation using a linear function). To clearly demonstrate the procedure, let $p\left(x_{i}, y_{i}\right)$ represent a pixel sampling from $m,$ where $x_{i}$, $y_{i}$ denotes as the coordinates of the corresponding pixel. Then the affine transformation can be expressed as:
\begin{equation}
\label{Eq.1}
A_{\theta}(p)=\left[\begin{array}{lll}
\theta_{11} & \theta_{12} & \theta_{13} \\
\theta_{21} & \theta_{22} & \theta_{23}
\end{array}\right] \cdot\left[\begin{array}{l}
x_{i} \\
y_{i} \\
1
\end{array}\right]
\end{equation}
where $\theta$ represents the parameters that determine the linear transformation. We pre-train a shallow regression network to predict those parameters. With the obtained parameters, STN can perform the affine transformation automatically without human involvement to roughly align the moving images $M$ to corresponding fixed images $F$. This regress network is called the affine transformation network (ATNet) in our framework. With ATNet, we can acquire the affine transformed predictions of the original moving images, which are represented as $M_{A} \supset\left\{m_{A}^{1}, m_{A}^{2} \cdots m_{A}^{k}\right\}$. These predictions are roughly aligned to $F$, and dense deformation transformations are needed to align the detailed local structures. It can be seen that only performing a linear transformation will not be able to capture the subtle differences between multi-contrast images. Besides, since affine transformations are global information-driven, the performance may be compromised when registered in low signals area. Therefore, predictions of the affine transformation network are treated as coarse registration images, which need to be further improved.

\subsection{Deformable transformation network – DTNet}
Deformable transformations are important for fine image registration. VoxelMorph \cite{balakrishnan2018unsupervised,dalca2018unsupervised,dalca2019unsupervised,balakrishnan2019voxelmorph} constructs a differentiable operation, which can be optimized through network training, on each pixel to realize image registration. Let us define $\varphi$ as the obtained transformation field. Each value in $\varphi$ represents an offset distance. Symbol $\circ$ refers to the transformation operator for $m^{k}$, which consists of pixel shifting and interpolation. For each pixel $p$ in $m^{k}$ transform to $p^{\prime}$ can be defined as:
\begin{equation}
\label{Eq.2}
p^{\prime}=p+\varphi(p)
\end{equation}
VoxelMorph performs an additional linear interpolation in neighboring pixels after the pixel transformation to avoid discontinuities in transformed images:
\begin{equation}
\label{Eq.3}
m \circ \varphi(p)=\sum_{q \in Z(p)} m(q) \prod_{dim \in{x, y,z}}(1-|p_{dim}-q_{dim}|)
\end{equation}
where $Z$ represents the regions composed of adjacent pixels. Through this differentiable interpolation operation, the predicted results are smoother and more realistic.

We employ VoxelMorph as our deformable transformation network (DTNet) to conduct fine image registrations. Some changes in the network architecture were adopted. For example, we adopted a deeper convolution structure to fully extract features. In addition, the activate function of ReLu is replaced by Leaky ReLu. More details about the architecture will be arranged in section IV.

\subsection{Coarse-to-fine multi-contrast image registration framework}
To reduce the challenges of unsupervised multi-contrasts image transformation, we propose a coarse-to-fine image registration framework. Specifically, we embed the pre-trained ATNet $\mathrm{D}_{\theta}(F, M)$, with frozen parameters into DTNet. The affine transformed predictions $M_{A}$ can serve as the inputs to DTNet. In this way, DTNet receives images that were roughly aligned to the fixed images with decreased image discrepancies. Different from existing methods that conduct two-step registrations of using affine transformations as preprocessing and then refine the predictions, the proposed framework adopts an end-to-end approach that conducts those operations in one architecture. Compared with the existing registration method, our method does not need to iterate over affine or deformation transformations. Meanwhile, we can obtain the affine transformed predictions and deformable transformed predictions as side outputs of the framework.

\subsection{Dual consistency-constrained bi-directional image transformation}
Intuitively, the registration procedure should be symmetrical, which refers to the bi-directional transformations between the moving images and the fixed images. This assumption was first proposed in \cite{avants2008symmetric} with an Euler Lagrange equation for iterative optimization and achieved great success in medical image registration. Inspired by this work, we propose a bi-directional image transformation method. 

As defined in the previous section, $\varphi$ is the transformation field for the forward transformation of registering moving images to fixed images. However, to inverse the transformation and restore the moving images, simply apply $-\varphi$ to the predictions of the forward deformation will not work because the correspondence between $\varphi$ and the image pixels has been destroyed by the forward shift. Let $\varphi_{i,j}$ be the displacement of the pixel $(i, j)$ in the moving image. After the forward transformation, pixel $(i, j)$ becomes pixel $(i^{'}, j^{'})$ in the registered image. Then, $-\varphi_{i,j}$ should be the inverse displacement of the pixel $(i^{'}, j^{'})$ instead of $(i, j)$, and we need to find the correspondence between $-\varphi_{i,j}$ and $(i, j)$. Accordingly, we constructed the inverse deformation field $\varphi^{-1}$ to guide the backward transformation. Instead of building a new network to generate $\varphi^{-1}$ from scratch \cite{wang2020lt} or integrate the negative velocity field \cite{dalca2019unsupervised}, we extrapolate $\varphi^{-1}$ from $\varphi$ to reduce complicated operations. Specifically, as $\varphi$ consists of the horizontal and vertical offsets in the 2D space, we first decompose $\varphi$ to obtain the two offset fields $\varphi_x$ and $\varphi_y$, respectively. Then, we warp the offset fields with the original $\varphi$ to form the deformed offset fields. By recombining the deformed offset fields, a new transformation field is generated. Finally, the inverse transformation field $\varphi^{-1}$ is obtained by multiplying with -1. In this way, we successfully align the transformation field with the pixels in the registered images. To sum up, the whole process can be represented by the following equation:
\begin{equation}
\label{Eq.4}
\mathrm\varphi^{-1}=-\Sigma_{x,y}\left(\varphi_{i} \circ \varphi\right)
\end{equation}

Since there are no reference images to evaluate the accuracy of the multi-contrast registration predictions, it is difficult to conduct the bi-directional registrations simultaneously from $M$ to $F$ and from $F$ to $M$. To combat this issue, we come up with a compromised solution that transfers the multi-contrast bi-directional image registration task to a unimodal image registration task, i.e. we use the predictions $M_{D} \supset\left\{m_{d}^{1}, m_{d}^{2} \cdots m_{d}^{k}\right\}$ instead of the fixed images $F$ to calculate the inverse transformed images: ${M}_{D}^{-1}=\mathrm{M}_{D} \circ \varphi^{-1}$. Here, we assume that $M_{D}^{-1}$ should still maintain the same distribution as $M_A$. Base on this, we use a consistency loss to accurate constraint $M_{D}^{-1}$ to $M_A$, which can be MSE or NCC. We can then obtain our integrated framework, the coarse-to-fine multi-contrast image registration framework with dual consistency constraint.

\subsection{Coarse-to-fine multi-contrast image registration framework with dual consistency constraint}

\begin{figure}[htbp]
\centering
\includegraphics[width=1\linewidth]{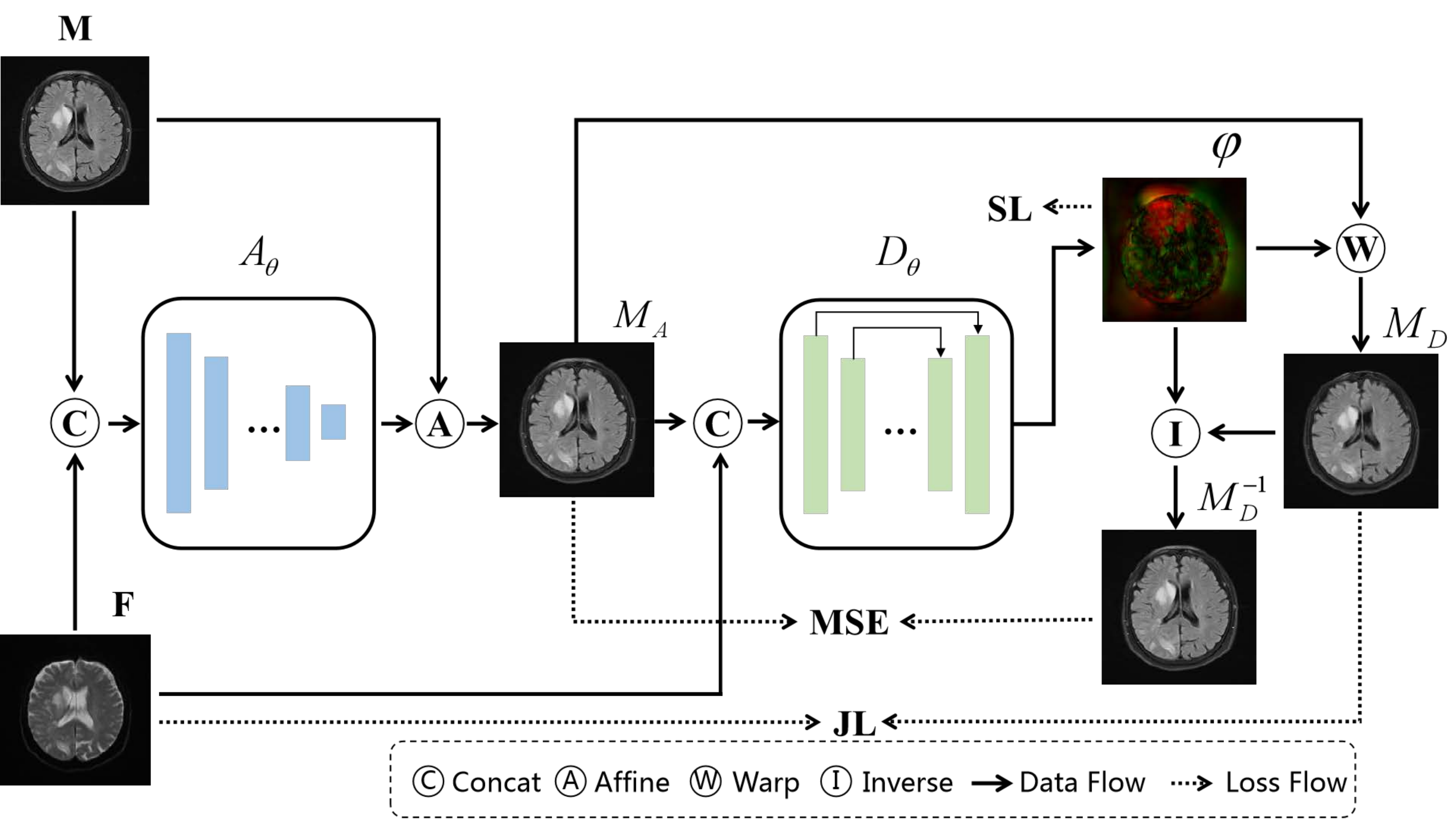}
\caption{The proposed coarse-to-fine multi-contrast image registration framework with dual consistency constraint. $A_{\theta}$ represents the pre-trained ATNet. $D_{\theta}$ refers to the DTNet. $M_D$ and $M_{D}^{-1}$ construct the bi-directional registration cycle.}
\label{Fig.2}
\end{figure}

Our coarse-to-fine multi-contrast image registration framework with dual consistency constraint is illustrated in Fig. \ref{Fig.2}. The framework consists of three main parts: 1) The pre-trained affine transformation network ATNet ($A_{\theta}$) for coarse affine registration. The input to ATNet is a pair of $M$ and $F$ MR multi-contrast images. The output is the affine transformation for coarse alignment from $M$ to $F$. The coarsely aligned images $M_A$ are the inputs to the subsequent deformable transformation network. It is important to note that once the pre-training is finished, the parameters of ATNet are frozen and no longer updated. 2) The deformable transformation network DTNet is to generate the final predictions. The input to DTNet is a concatenation of $F$ and $M_A$. The output is a densely transformation field $\varphi$. With $\varphi$, the final prediction $M_D$ is generated. 3) A dual consistency constraint. We propose a novel inverse transformation from $M_D$ to $M_{D}^{-1}$ to further enhance the registration performance. We calculated the inverse transformation field $\varphi^{-1}$ and warp $M_D$ with it to obtain $M_{D}^{-1}$. By enforcing a similarity measure between $M_{D}^{-1}$ and $M_A$, we achieve the dual consistency constraint. With the bi-directional registration strategy, undesirable interpolation during image registration is expected to be suppressed and a more accurate registration can be obtained.

\subsection{Loss function}
As indicated in Fig. \ref{Fig.2}, multiple loss functions are utilized to optimize the multi-contrast MR image registration framework. For simplicity, we use $\xi_{\theta}(\cdot)$ to represent an undefined network which can be either ATNet ($A_{\theta}$) or DTNet ($D_{\theta}$).

The most important loss function used is Mutual Information (MI), which can measure the distribution dependence between two random variables [45]. Here, we define two marginal probability distributions, $p_{F}(f)$ and $p_{M}(m)$, and a joint probability distribution $p_{F,M}(f,m)$. MI measures the degree of dependence between $F$ and $M$ by calculating the distance between the joint distribution $p_{F,M}(f,m)$ and the distribution $p_{F}(f)p_{M}(m)$ by means of the Kullback-Leibler measurement\cite{vajda1989theory}. MI loss ($MI$) can be written as Eq. 5:

\begin{equation}
\label{Eq.5}
\begin{aligned}
&MI(F, M)= -\iint p_{F,M}(f, m) \log (\frac{p_{F,M}(f, m)}{p_{F}(f)p_{M}(m)})dxdy
\end{aligned}
\end{equation}

If $F$ and $M$ are independent, $p_{F,M}(f, m)$ is equal to $p_{F}(f)p_{M}(m)$, and $MI(F, M)$ will be zero, which means that there is no mutual information between the two variables. Maximization of MI is a general and powerful criterion because no assumptions are made regarding the nature of this dependence and no limiting constraints are imposed on the image content of different modalities involved \cite{maes1997multimodality}. 

Since MR images are usually in grayscale with background values close to 0, we suggest no signals should appear in the background regions of registered images. Based on this, we propose a prior knowledge-based background suppressing loss function: $MSE(f,m)=(f-m)^2$ when $f$ are background pixels.

Combing the MI loss function and the prior knowledge-based background suppressing loss function, we obtain the first loss function, which is called a prior knowledge-based joint loss function ($\mathrm{JL}\left(F, \xi_{\theta}(F, M), \lambda\right)$) as shown in Eq.6:
\begin{equation}
\begin{aligned}
\label{Eq.6}
\mathrm{JL}(F, \xi_{\theta}(F, &M), \alpha, \beta)\quad=\\&\sum_{f, m}(\alpha M I(f, \xi_{\theta}(f, m))+\\
&\beta \sum_{i}\left\{\begin{array}{c}
MSE(f_{i}, \xi_{\theta}(f, m)_{i}), \quad \text { if } \quad f_{i}<\gamma \\ 0, \quad \text { otherwise }
\end{array}\right.
\end{aligned}
\end{equation}
where $i \in N$ represents the pixels in images, $\gamma$ is a threshold obtained from the data set to determine whether the pixel is background or not, $\alpha$ and $\beta$ are adjust factors to balance the two losses. JL can not only constrain the global image alignment by maximizing MI, but also penalize the incorrect predictions in defined regions. This makes the predictions more in line with the nature of medical images.

The second loss function we use is to meet the dual consistency constraint. A simple MSE loss is calculated instead of MI loss between $M_{D}^{-1}$ and $M_{A}$. The utilization of MSE loss is not fixed and can be replaced by similar losses, such as NCC or $L_{1}$-norm.

The last loss function is calculated to constrain the transformation field $\varphi$. Transformation may occur with an irregular displacement without constraint, whereas the above mentioned two losses can still be small through the interpolation algorithm. To prevent such situations, a spatially smooth loss function is designed to refine the transformation field $\varphi$:
\begin{equation}
\label{Eq.7}
\mathrm{SL}(\varphi)=\sum_{f, m}|\nabla \varphi(f, m)|^{2}
\end{equation}
where $\nabla(\cdot)$ represent the calculation of gradients. By limiting the gradient of the deformation field, we make sure that the transformation field is smooth, and extreme pixel displacement can be avoided.

The overall loss function to optimize the framework is calculated as shown in 
\begin{equation}
\begin{aligned}
\operatorname{Loss}_{total}(F, M)=&\lambda_{1} S L(\varphi)+J L(F, D_{\theta}(F, M), \lambda_{2},\lambda_{3})+\\
&\lambda_{4} M S E(\mathrm{A}_{\theta}(F, M), \mathrm{D}_{\theta}^{-1}(F, M))
\end{aligned}
\label{Eq.8}
\end{equation}
The equation contains four adjust factors $\lambda_{i \in\{1,2,3,4\}}$. These are hyper-parameters that can be set to different values according to the experiment.
\section{Experiments and Results}
\label{sec:experiments and results}
In this section, we verify the effectiveness of the proposed methods through extensive experiments. In clinical practices, FLAIR and DWI are the most commonly used MR weighted sequences. Thus, our image registration experiments are mainly conducted with FLAIR and DWI data.

\subsection{Dataset}
The multi-contrast MR data were collected by Guizhou Provincial People's Hospital. This retrospective study was approved by the institutional review board of the hospital with the written informed consent requirement waived. All patient records were de-identified before analysis and reviewed by the institutional review boards to guarantee no potential risk to patients. The researchers who conduct the registration tasks have no link to the patients to prevent any possible breach of confidentiality.

In total, data from 555 patients are utilized with or without stroke lesions. Each patient was scanned with five sequences: T1 weighted, T2 weighted, FLAIR, ADC, and DWI. All images were obtained with a Siemens 1.5T scanner. Of the 555 cases, 466 are provided with the scanning information while the others are not. Two sets of experiments were conducted, one with the physical space alignment according to the provided scanning information and the other without. In the first set of experiments, only the 466 cases with scanning information were utilized. 426 cases were randomly selected as the training set and the remaining 40 cases as the test set. Stroke lesions in DWI and FLAIR images of the test set were annotated by experienced clinicians for quantitative result evaluation. In the second set of experiments, all 555 cases were used. Since no physical space pre-alignment was performed, the 89 cases without scanning information were included in the training set. All the data are resized to $224 \times 224$ with intensity normalized $[0, 1]$.

\subsection{Implementation details}
Theoretically, ATNet and DTNet can adopt various network structures. In this study, we prefer simple network structures to reduce computational complexity. We will show in the results section that even with the selected simple network structures, our proposed method can still achieve good registration performance.

ATNet is implemented with a regression network, which contains five downsampling blocks and two fully-connected layers. Each downsampling block consists of two $3 \times 3$ convolutional layers followed by a $2 \times 2$ max pooling layer. The convolution operation is always followed by batch normalization and leaky ReLU activation unless otherwise specified. Finally, two fully-connected layers is appended to generate the 6 transformation parameters. With these parameters, affine transformations are performed. The channels of the downsampling blocks and the last two fully-connected layers are set as 16, 32, 32, 64, 64, 128, 32, and 6, respectively. ATNet has about 589k trainable parameters.

DTNet is modified from the famous UNet with an encoder-decoder architecture \cite{ronneberger2015u}. The encoder of DTNet is the same as the above mentioned ATNet, whereas the decoder is designed symmetrically to the encoder. For the last layer, we utilized two $3 \times 3$ convolutions with linear activations and then, the final transformation field $\varphi$ can be obtained. DTNet has about 1478k trainable parameters.

Multiple comparison methods are adopted, including VoxelMorph (VM) \cite{balakrishnan2019voxelmorph}, VoxelMorph-diff (VM-diff) \cite{dalca2019unsupervised}, LT-Net \cite{wang2020lt}, and Symmetric Normalization (SyN) \cite{avants2008symmetric}. VM is the most famous deep learning-based registration algorithms developed in recent years. We slightly adjust the method (using the MI loss) to make it suitable for multi-contrast image registration. For LT-Net, we discarded the label transfer part and only kept the main registration framework with the inverse module. SyN is a top-performing brain registration algorithm. It is implemented in the publicly available Advanced Normalization Tools (ANTs) software package \cite{avants2011reproducible} with a MI constraint for multi-contrast MR image registration. In our implementation, SyN has two designs: 1) Moving images go through ANTs-based affine transformations and SyN, represented as ‘SyN(Affine)’; 2) Moving images go through SyN only, represented as ‘SyN(Only)’. Since the GPU implementations for these two methods are not currently available, CPU implementations are utilized and the registration speed is reported accordingly.

Our method is implemented using Keras with a Tensorflow backend on a NVIDIA Titan Xp GPU. All experiments are based on 2D slices. During training, data augmentation methods are applied including random translations, rotations, dilations, and horizontal flip. The batch size is set to 32, and the learning rate is set to 0.01 with an Adam optimizer. Pre-training ATNet takes about 25 minutes, and the entire framework including DTNet requires another 20 minutes to optimize. The four weights in the loss function, $\lambda_{i \in\{1,2,3,4\}}$, were set to {1, 4, 100, 100} empirically. The threshold factors $\gamma$ in the JL was set to 0.1. Our code will be available online at \url{https://github.com/SZUHvern/TMI_multi-contrast-registration}.

\begin{figure*}[htbp]
\centering
\includegraphics[width=1\linewidth]{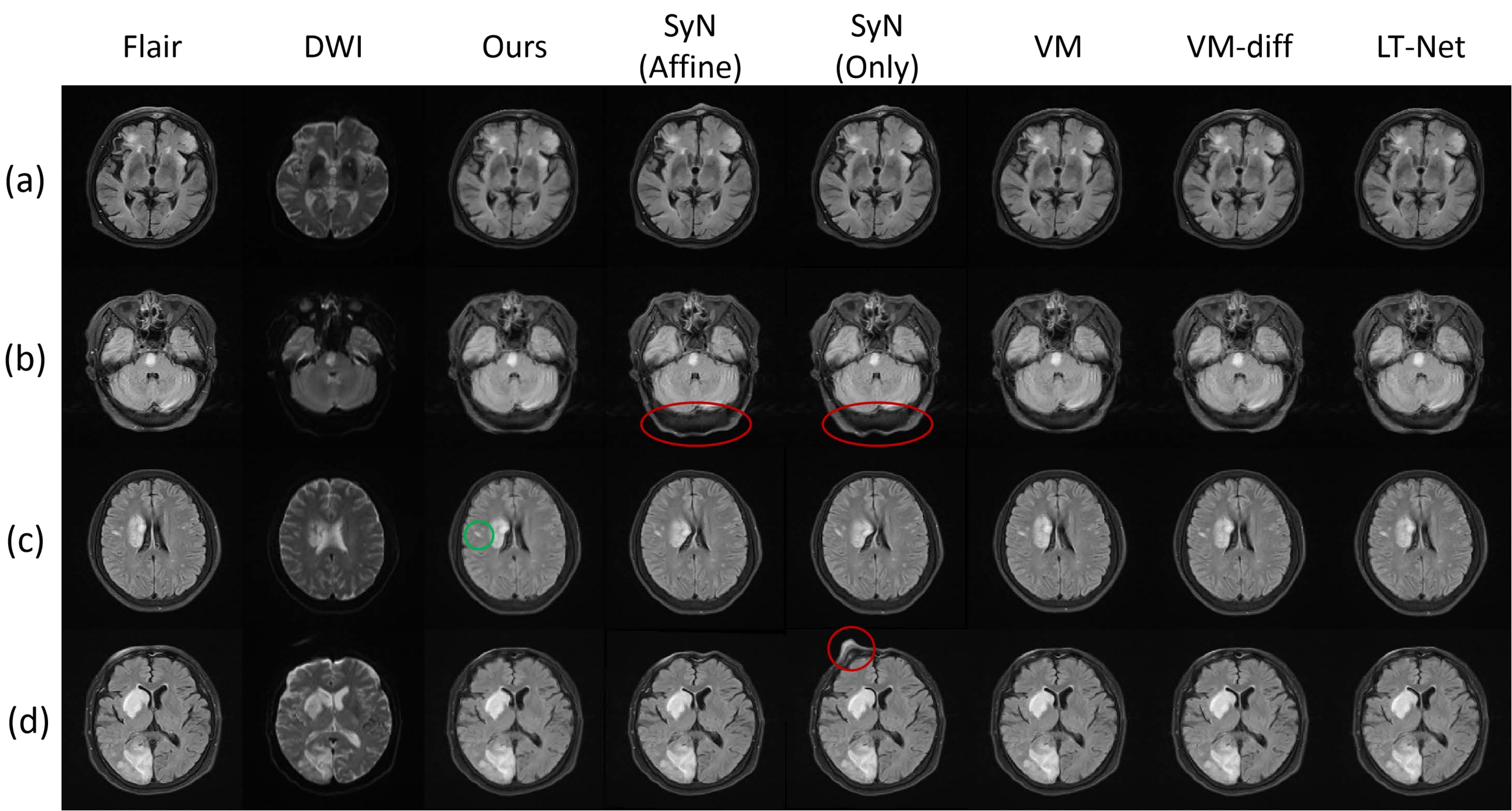}
\caption{Qualitative registration results of different methods. The green circle denotes regions that our method registered better than other methods while the red circle denotes unexpected predictions based on the specific comparison methods.}
\label{Fig.3}
\end{figure*}

\subsection{Results of multi-contrast MR image registration}
In this section, qualitative and quantitative image registration results are reported. Quantitative results are calculated with regard to the alignment of stroke lesions between registered moving images and fixed images. Please note that there is still a lack of measurement metrics to characterize multi-contrast MR image registration. Although the area or shape of the stroke lesions may be differently presented in multi-contrast images, we believe that alignment between the stroke lesions can still reflect the registration performance.

We evaluated our method using Dice, Recall, and Precision, which are commonly used in computer vision. They are important indicators to assess the overall difference between our predictions and the ground truths. The exact formulas to calculate the three scores are: $Dice = 2TP/(2TP+FP+FN)$, $Recall = TP/(TP+FN)$, and $Precision = TP/(TP + FP)$. Here, TP (true positive) indicates the numbers of correctly register pixels, FP (false positive) indicates the numbers of pixels that the model register negative as positive, and FN (false negative) indicates the numbers of pixels that the model register positive as negative. We calculate these scores based for each case individually and report the average results. In addition, in order to quantify the deformation regularity, we calculate the Jacobian determinant $J_\varphi$ as the derivative of the deformation field, and $|J_\varphi|<0$ indicates the locations where folding has occurred. We report the number of pixels where $|J_\varphi|<0$.

Example predictions of different methods are shown in Fig. \ref{Fig.3}. For example (a), the comparison methods generated deformed skulls while our method can keep the structure very well. In regions with low signals, such as example (b), iterative methods, SyN (Affine) and SyN (Only), show unexpected deformations and the tissues look abnormal. In example (c), the deep learning-based comparison methods cannot fully register the stroke lesions. Our method can still perform well thanks to the more robust registration flow we designed. Finally, when scanning artifacts exist in remote regions (example (d)), SyN (Only) shows obvious image distortion in order to fit the artifacts. Overall, satisfactory results are achieved by all the registration methods, and our method performs especially well for challenging cases with artifacts, sharp changes, etc.

\begin{table*}
\caption{Quantitative measurement of the stroke region registration results and the required test time.}\label{Table.1}
\centering
\begin{tabular}{cccccc}
\toprule
\multicolumn{1}{c}{Method} & \multicolumn{1}{c}{Dice} & \multicolumn{1}{c}{Precision} & \multicolumn{1}{c}{Recall} & \multicolumn{1}{c}{\begin{tabular}[c]{@{}c@{}}Sec/Slice\\ (GPU)\end{tabular}} & \multicolumn{1}{c}{\begin{tabular}[c]{@{}c@{}}Sec/Slice\\ (CPU)\end{tabular}} \\
\hline
Undef & $0.7822 \pm 0.0974$ & $0.8491 \pm 0.1007$ & $0.7326 \pm 0.1175$ & - & - \\
SyN(Only) & $0.8101 \pm 0.0979$ & $0.8734 \pm 0.1137$ & $0.7669 \pm 0.1237$& - & $1.4446$ \\
SyN(Affine) & $ 0.8157 \pm 0.0950$ & $0.8769 \pm 0.1061$ & $0.7736 \pm 0.1244$& - & $2.0335$ \\
LT-Net & $ 0.7960 \pm 0.1043$ & $0.8480 \pm 0.1087$ & $0.7595 \pm 0.1220$& $0.0162$ & $0.2156$ \\
VM-diff& $ 0.8011 \pm 0.0991$ & $0.8305 \pm 0.1059$ & $0.7826 \pm 0.1184$& $0.0405$ & $0.1801$ \\
VM & $0.8053 \pm 0.0954$ & $\bm{0.8861 \pm 0.0856}$ & $0.7534 \pm 0.1322$ & $0.0109$ & $0.1699$ \\
ATNet & $0.8067 \pm 0.0935$ & $0.8699 \pm 0.0948$ & $0.7590 \pm 0.1128$ & $\bm{0.0100}$ & $\bm{0.0299}$ \\
Ours & $\bm{0.8397 \pm 0.0756}$ & $0.8856 \pm 0.0808$ & $\bm{0.8081 \pm 0.1069}$ 
& $0.0223$ & $0.2037$ \\ \bottomrule
\end{tabular}
\end{table*}

The quantitative results are listed in Table I. Without registration, stroke lesion annotations in FLAIR and DWI images are misaligned with an average Dice score of 0.7822, which reflects the need for multi-contrast image registration. ATNet gets 0.8067, reflecting that even with the physical space alignment, linear transformations is still needed to achieve accurate registration. The methods without the affine transformation, SyN (Only) generate a similar result of 0.8101. It is improved to 0.8157 by introducing the affine transformation (SyN (Affine)). Our method achieves the highest score of 0.8397, proving its effectiveness in handling the multi-contrast image registration problem.

Efficiencies of the different methods are also compared. For fair comparisons, all the methods are tested on a CPU. SyN (Affine) is the least efficient method that spends 2.0335 seconds to register one image slice, and ATNet has the highest efficiency which needs only 0.02 seconds. Comparing with the most competitive method, SyN (Affine), our method is about 10 times faster with better registration results. It can achieve the registration of one 3D image case (20 slices) within 5 seconds, which is sufficient for real-time diagnosis in clinical practices. The time spent can be further shortened to within 0.5 second/case when testing on a GPU.

\subsection{Visualization of the transformation field}

\begin{figure}[htbp]
\centering
\includegraphics[width=1\linewidth]{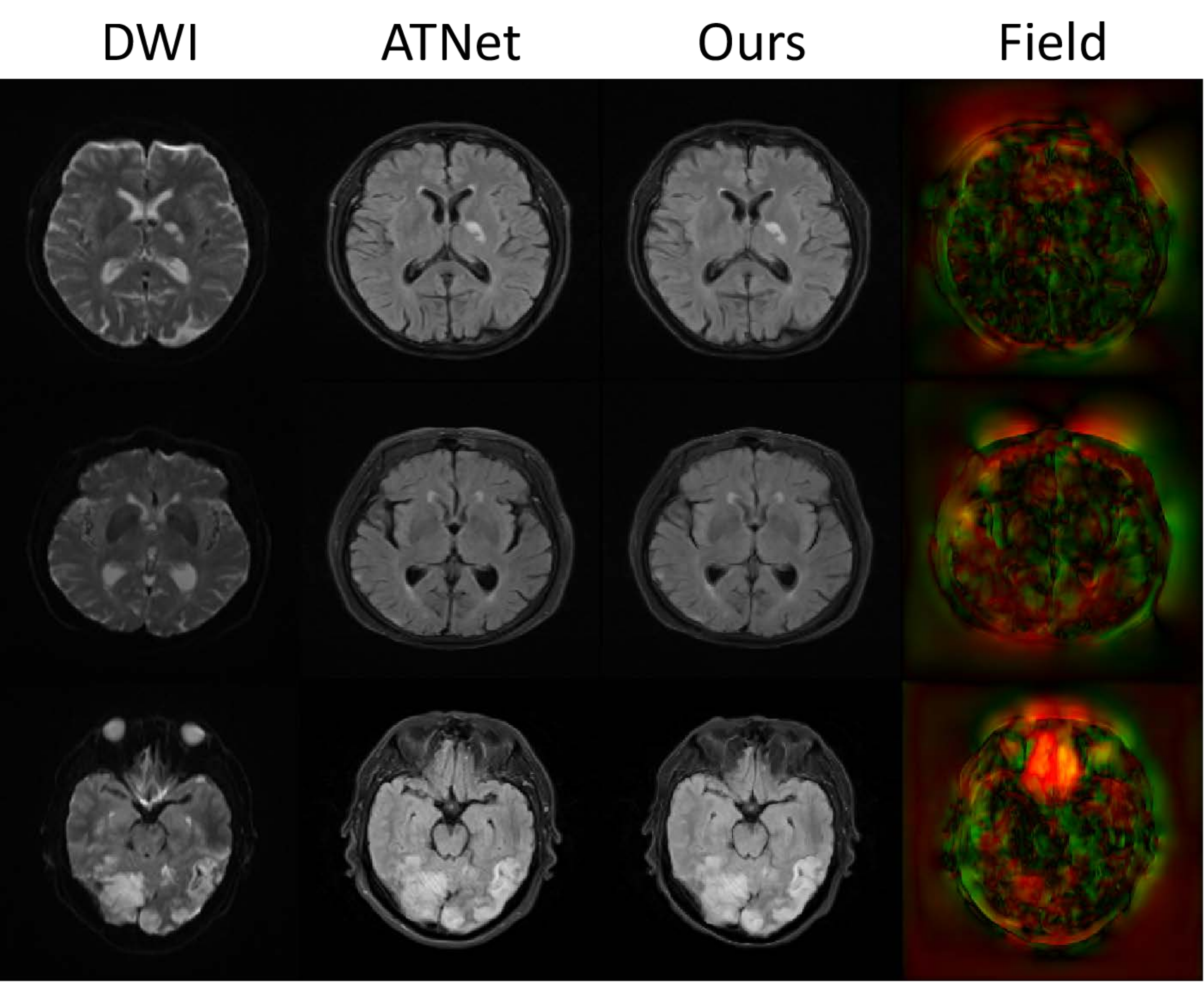}
\caption{Visualizations of the transformation fields $\varphi$. Red color indicates the transformation in the horizontal direction and green indicates the transformation in the vertical direction. Higher red or green color signals indicate larger transformations.}
\label{Fig.4}
\end{figure}

Visualizations of example transformation fields $\varphi$ are shown in Fig. \ref{Fig.4}. These examples indicate that even after the physical alignment and affine transformation, large deformations (indicated by the red and green signals in the transformation fields) are still needed for accurate registrations. As a result, physical alignment, affine transformation, and deformable transformation, especially the latter two, are simultaneously required in applications.

\subsection{Time consumption analysis of inverse transformation}
To investigate the efficiency of the proposed inverse transformation, we compare the time consumption of our proposed method with the existing inverse methods, VM-diff \cite{dalca2019unsupervised} and LT-Net \cite{wang2020lt}.

VM-diff \cite{dalca2019unsupervised} introduced an inverse deformation by adding a differential and integral layer (to generate velocity field) combined with a spatial transformation layer. The inverse deformation field is then obtained by iterating the negative velocity field. Specifically, this method split the registration into T=7 integration steps and then warp moving images according to the computed diffeomorphic field $\varphi^{-1}$ using a spatial transform layer. Comparing with our method of calculating the inverse deformation field in one step, this method requires more operation steps. 

LT-Net \cite{wang2020lt} is a cycle-correspondence learning method for atlas-based segmentation. This method builds a new network to learn the inverse deformable field and achieves the inverse transformation through a transformation layer. Since the inverse deformation field is realized by a new network, this method is more complicated than ours.

We conduct experiments to quantitatively compare the time consumption of different methods to prove our analysis (Table \ref{Table.2}). For fair comparisons, we implemented all methods with the same neural networks (DTNet) except for the inverse operation, and thus, the obtained time consumptions are different from those shown in Table \ref{Table.1}. As expected, our method is the fastest with a registration speed of 0.0233 s/slice on a GPU.

\begin{table}[htbp!]
\centering
\caption{Quantitative time comparison between different inverse methods.}
\label{Table.2}
\setlength{\tabcolsep}{1mm}{
\begin{tabular}{ccc}
\toprule
Method                 &Sec/Slice(GPU) & Sec/Slice(CPU)\\
\hline
VM-diff                & $0.0565$ & $0.2147$  \\
LT-Net                 & $0.0320$ & $0.2479$ \\
Ours                   & $\bm{0.0223}$ &$\bm{0.2037}$\\ 
\bottomrule
\end{tabular}}
\end{table}

\subsection{Ablation experiment}
We also conducted extensive ablation experiments to verify the effectiveness of the proposed framework. Firstly, we investigate the influence of network widths on the registration performance under two learning rates. Then, we inspect the importance of the all the proposed structures. Finally, we discussed the influence of JL's parameter selection on the prediction result. 

\begin{figure}[htbp]
\centering
\includegraphics[width=1\linewidth]{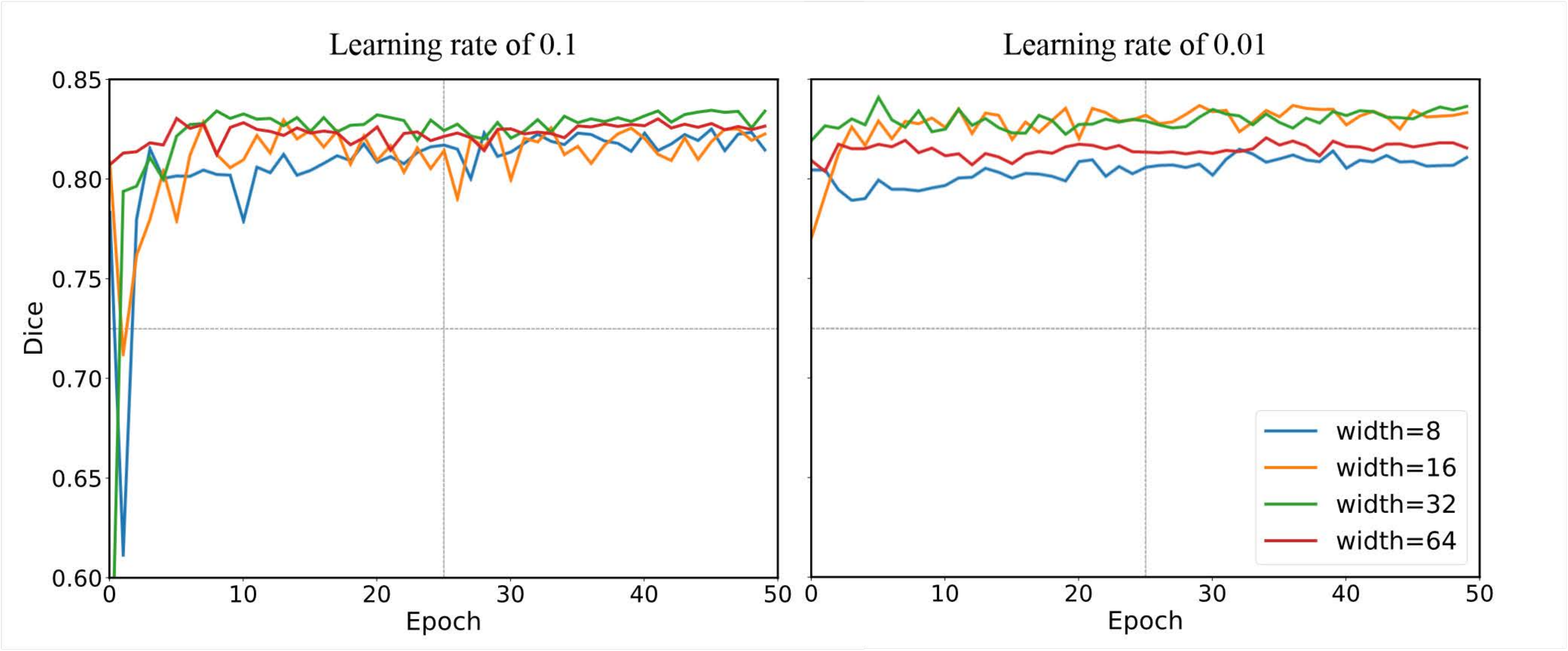}
\caption{Results of networks with different widths (8, 16, 32, 64) under two learning rates of 0.1 and 0.01. The width value represents the number of feature maps in the first block of DTNet.}
\label{Fig.5}
\end{figure}

In Fig. \ref{Fig.5}, we show the Dice scores of networks with different widths under two learning rates. Although the larger learning rate can lead to relatively faster convergence, fluctuated Dice score curves indicate that the training is unstable. Especially for width of 32 network, a smaller learning rate might be more appropriate. For the different network widths, significantly worse performance is observed with a width of 8 and 16, which might indicate that the network is not able to capture the complex image properties. Wider networks with widths of 32 and 64 show similar performance and the network with a width of 32 performs slightly better. It is worth noting that there is no overfitting in all implementations, which indirectly proves the suitability of our method for the multi-contrast image registration task.

To inspect the importance of the proposed structures, we conducted experiments progressively under different settings. The results are listed in Table \ref{Table.3}. DTNet performed well when compared with the linear transformation ATNet, showing the advantages of deformable transformation. Besides, when these two types of methods are used in combination, the performance is further improved, achieving a Dice score of 0.8258. These experiments prove that the proposed coarse-to-fine framework is effective. Utilizing this framework, we add the proposed JL constraint, and the Dice score is increased by another $0.8\%$ thanks to the effective suppression of the wrongly predicted background pixels. Our final model is constructed by introducing the proposed dual consistency constraint, achieving the best Dice score of 0.8397.

We derived the Jacobian determinant to calculate the number of folding pixels (the lower the better) to check the model effectiveness. The coarse-to-fine framework, ATNet + DTNet, gets a number of $30\pm30$, which is significantly lower than that of DTNet ($46\pm50$). However, this number is slightly increased when adding the JL constraint. We suspect that JL is designed for background error suppression, which might lead to the folding of unexpected background pixels. Nevertheless, with the introduction of the dual consistency constraint, the number is largely reduced to $13\pm15$. This proves that the dual consistency constraint can effectively suppress the occurrence of pixels folding through the inverse transformation.

\begin{table*}[htbp]
\centering
\caption{Quantitative results comparison between different methods.}
\label{Table.3}
\setlength{\tabcolsep}{1mm}{
\begin{tabular}{ccccc}
\toprule
Method         & Dice              & Precision         & Recall           & $|J_\varphi|<0$ \\
\hline
Undef          & $0.7822 \pm 0.0974$ & $0.8491 \pm 0.1007$ & $0.7326 \pm 0.1175$ & $-$ \\
ATNet          & $0.8067 \pm 0.0935$ & $0.8699 \pm 0.0948$ & $0.7590 \pm 0.1128$& $-$\\
DTNet          & $0.8117 \pm 0.0919$ & $0.8727 \pm 0.0979$ & $0.7690 \pm 0.1173$& $46 \pm 50$ \\
ATNet+DTNet    & $0.8258 \pm 0.0787$ & $0.8847 \pm 0.0826$ & $0.7905 \pm 0.1112$& $30 \pm 30$\\
ATNet+DTNet(JL)& $0.8339 \pm 0.0840$ & $\bm{0.8889 \pm 0.0801}$ & $0.7955 \pm 0.1165$& $60 \pm 50$\\
Ours           & $\bm{0.8397 \pm 0.0756}$ &$0.8856 \pm 0.0808$&$\bm{0.8081 \pm 0.1069}$ &$13 \pm 15$\\ 
\bottomrule
\end{tabular}}
\end{table*}

\begin{figure}[htbp]
\centering
\includegraphics[width=1\linewidth]{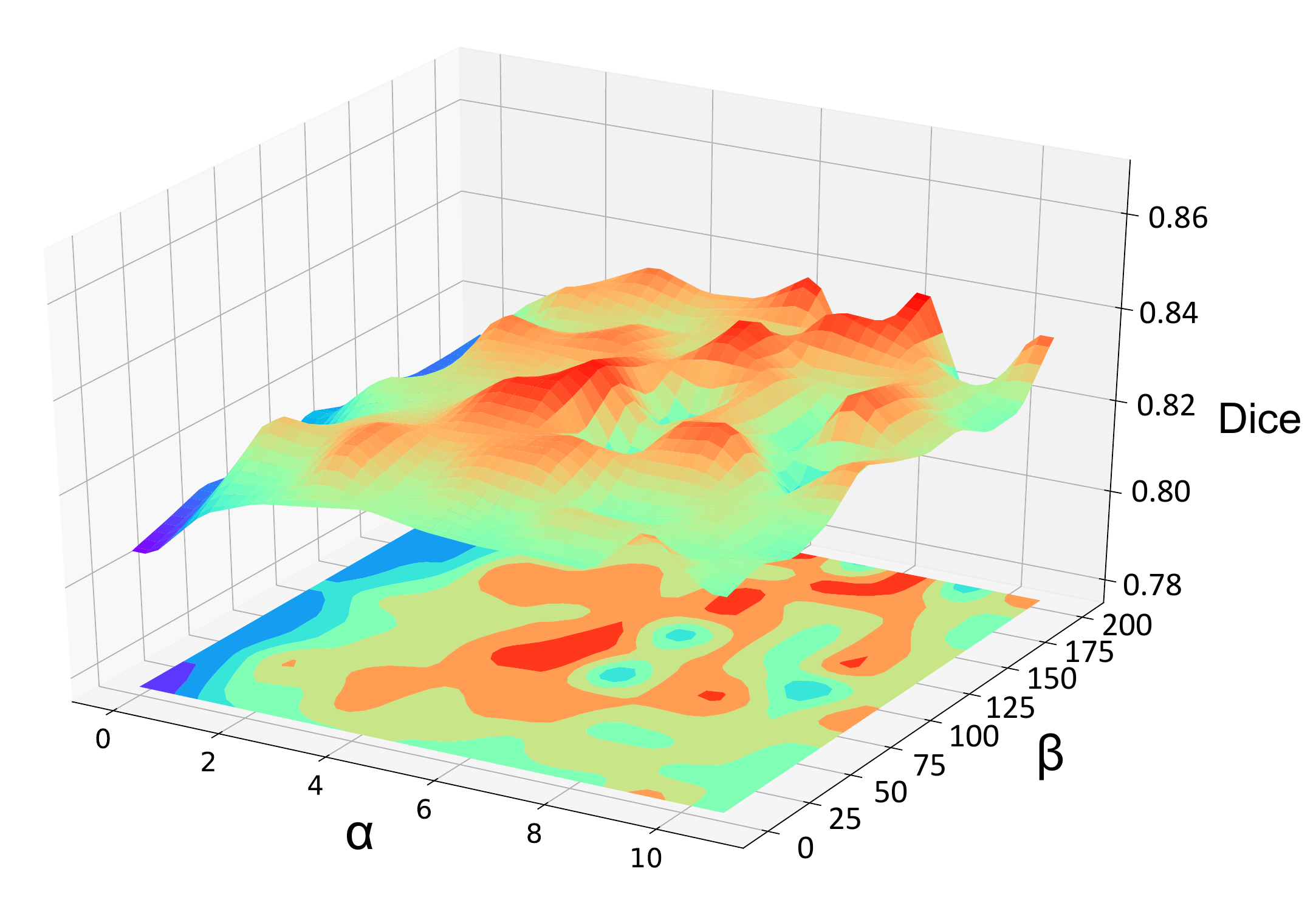}
\caption{Influence of the weights ($\alpha$ and $\beta$) in the proposed JL on the registration performance.}
\label{Fig.6}
\end{figure}
As stated in the previous sections, the four weights in the JL function (Eq. 8), $\lambda_{i \in\{1,2,3,4\}}$, were empirically set to {1, 4, 100, 100}. Here, we conducted experiments by fixing $\lambda_{1}$ and $\lambda_{4}$ to investigate the influence of $\lambda_{2}$ and $\lambda_{3}$, which are also the $\alpha$ and $\beta$ in Eq. 6 that control the relative contributions of MI loss and the prior knowledge-based background suppressing loss. In details, we checked different $\alpha$ values from 0 to 10 with a step size of 1, and different $\beta$ values from 0 to 200 with a step size of 20. The results are shown in Fig. \ref{Fig.6}. Two conclusions can be made. Firstly, with the increase of $\alpha$, the registration performance gradually improves until the Dice scores fluctuate around 0.83. This indicates that MI is important for accurate image registration. Secondly, with the increase of $\beta$, the registration performance also improves slightly. This confirms that the proposed prior knowledge-based background suppressing loss can help MI loss better optimize the network. The best Dice score of 0.8397 is achieved when $\alpha=4$ and $\beta=100$, which is much better than the Dice score of 0.8289 when $\alpha=4$ and $\beta=0$. Overall, the registration performance is quite robust with changing $\alpha$ and $\beta$ values, and the proposed JL is effective. These results in all confirm that the proposed coarse-to-fine architecture, JL, and the dual consistency constraint can successfully enhance the multi-contrast MR image registration performance.

\subsection{Experiment on data without scanning information}
There are occasions when the scanning information, including the pixel spacing and field of views, is lost. Without the scanning information, multi-contrast images cannot be pre-aligned in the physical space, which brings great difficulties to the accurate image registration task.
\begin{figure}[htbp]
\centering
\includegraphics[width=0.9\linewidth]{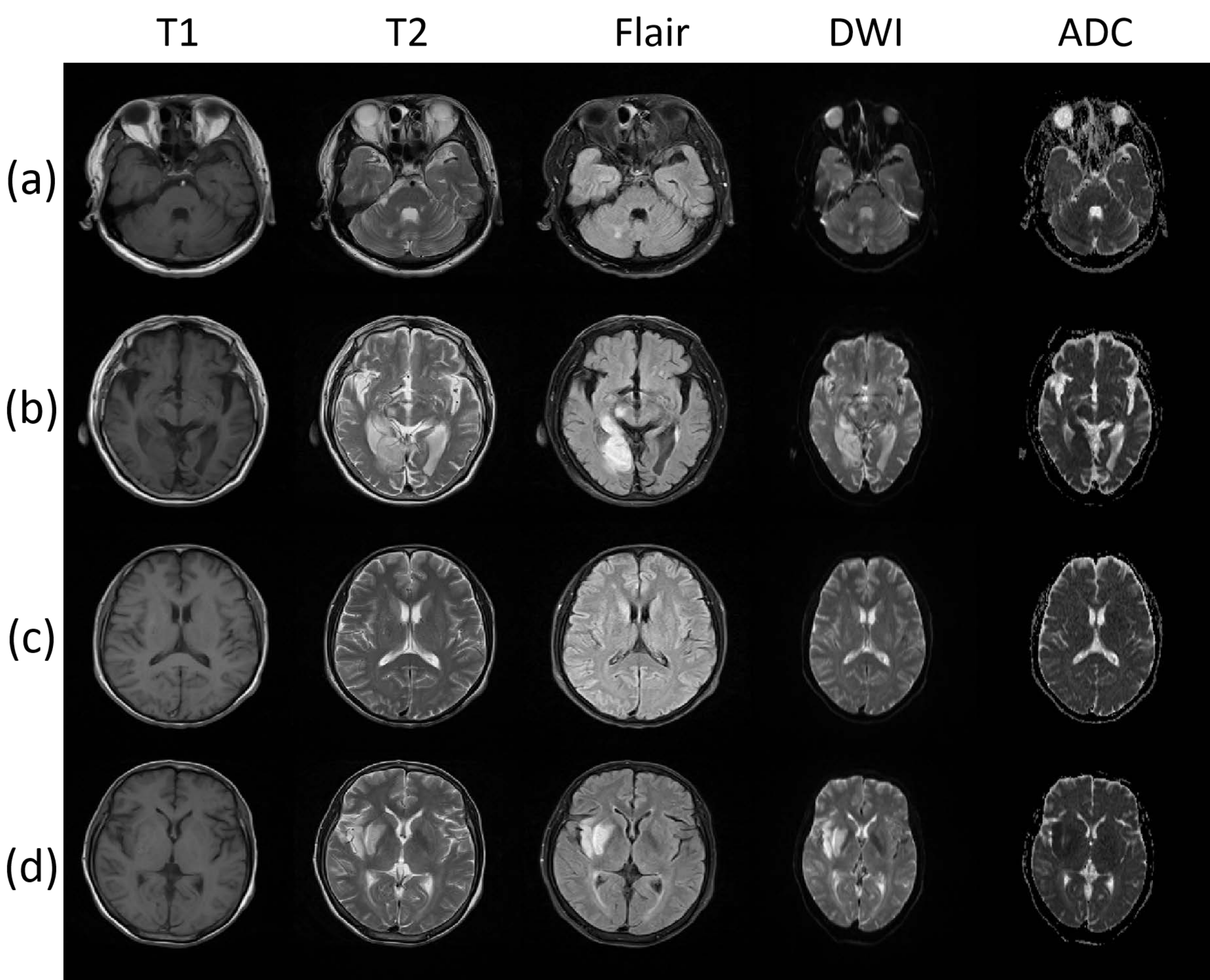}
\caption{MR multi-contrast brain images without physical alignment in advance. The images show large discrepancies between images acquired with different contrasts.}
\label{Fig.7}
\end{figure}
To increase the application capability of the proposed method, we conduct experiments without the first step physical space alignment. In this set of experiments, all the collected 555 image data are utilized. Some examples are shown in Fig. \ref{Fig.7}. It can be observed that due to the different imaging parameters, large discrepancies exist between the different contrast images.

The registration results of different methods are reported in Table \ref{Table.4}. Overall, worse results are obtained in these experiments compared to those achieved with the physical space alignment (Table \ref{Table.1}). Before any registration, stroke lesion annotations in FLAIR and DWI images are largely misaligned with an average Dice score of only 0.3472. Compared to the scores achieved with the first step physical space alignment (Table \ref{Table.1}), the performance of SyN (Only) is dramatically decreased by more than 20$\%$ (0.5880 vs. 0.8101). SyN (Affine) obtains a slightly decreased score of 0.8048. The scores of all the learning-based comparison methods are decreased by roughly 4$\%$. Our method still maintains a good performance with a score of 0.8260, which is only decreased by 1.37$\%$. It indicates that the proposed method generalizes well to difficult tasks, and thus, the robustness is improved. Considering the time complexities, our proposed method becomes better than the time-consuming iterative-based method of SyN (Affine). Overall, when facing more challenging tasks, our method can still maintain good registration performance with satisfactory registration speed.

\begin{table*}
\caption{Quantitative measurement of the stroke region registration results and the required test time based on data without pre-alignment.}\label{Table.4}
\centering
\begin{tabular}{cccccc}
\toprule
\multicolumn{1}{c}{Method} & \multicolumn{1}{c}{Dice} & \multicolumn{1}{c}{Precision} & \multicolumn{1}{c}{Recall} & \multicolumn{1}{c}{\begin{tabular}[c]{@{}c@{}}Sec/Slice\\ (GPU)\end{tabular}} & \multicolumn{1}{c}{\begin{tabular}[c]{@{}c@{}}Sec/Slice\\ (CPU)\end{tabular}} \\
\hline
Undef & $0.3472 \pm 0.2390$ & $0.3092 \pm 0.2131$ & $0.3999 \pm 0.2789$ & - & - \\

SyN(Only) & $0.5880 \pm 0.2502$ & $0.5539 \pm 0.2628$ & $0.6534 \pm 0.2584$ & - & 3.0238 \\
SyN(Affine) & $ 0.8084 \pm 0.0989$ & $\bm{0.8682 \pm 0.1411}$ & $0.7565 \pm 0.1510$ & - & 3.7044 \\
LT-Net & $ 0.7599 \pm 0.1371$ & $0.8283 \pm 0.1338$ & $0.7178 \pm 0.1597$& $0.0152 $& $0.2242$ \\
VM-diff& $ 0.7583 \pm 0.1181$ & $0.8166 \pm 0.1326$ & $0.7181 \pm 0.1283$& $0.0401$& $0.1776$ \\
VM & $0.7672 \pm 0.1281$ & $0.7984 \pm 0.1358$ & $0.7480 \pm 0.1403$ & 0.0109 & 0.1792 \\
ATNet & $0.7603 \pm 0.1279$ & $0.8066 \pm 0.1307$ & $0.7267 \pm 0.1438$ & $\bm{0.0101}$ & $\bm{0.0351}$ \\
Ours & $\bm{0.8260 \pm 0.0761}$ & $0.8666 \pm 0.0921$ & $\bm{0.7981 \pm 0.0989}$ & 0.0201 & 0.2176 \\ \bottomrule
\end{tabular}
\end{table*}

\begin{figure}[htbp]
\centering
\includegraphics[width=1\linewidth]{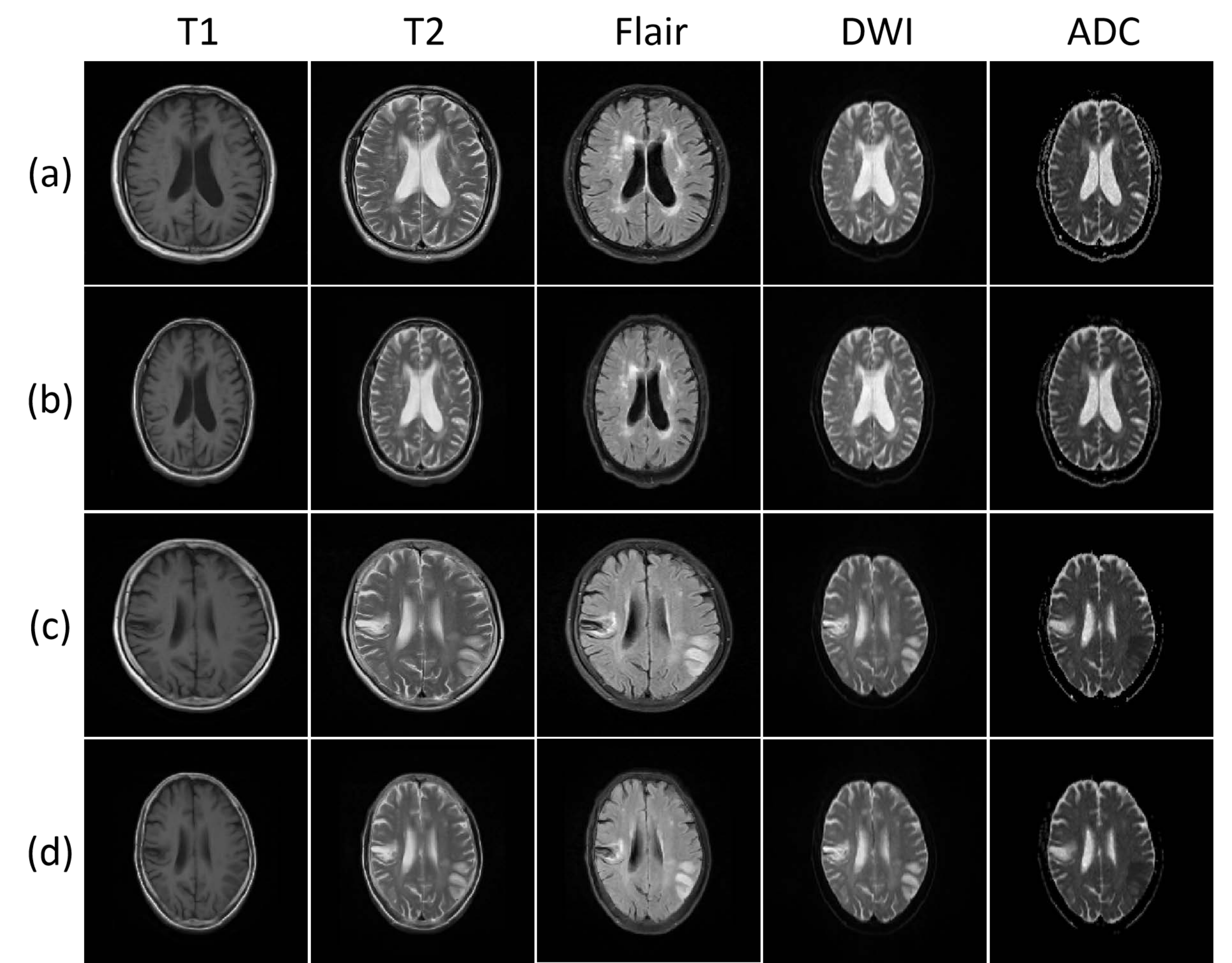}
\caption{Example results from registering MR images without pre-alignment acquired by the three structural sequences (T1 weighted, T2 weighted, and FLAIR) to DWI. (a) and (c) are two image slices selected from one patient without pre-alignment. (b) and (d) are the corresponding registration results.}
\label{Fig.8}
\end{figure}

Moreover, we also tested to the structural MR images acquired with the three contrasts (T1 weighted, T2 weighted, and FLAIR) to DWI images using the proposed method (Fig. \ref{Fig.8}). Results indicate that our method can also perform quite well, which shows the general applicability of our method when handling different multi-contrast MR image registration tasks. It again validates that robustness of our method, and its high potential to be applied in clinical practices.

\section{Conclusion}
\label{sec: conclusion}
Multi-contrast MR image registration is critical for many clinical applications. Existing registration methods are limited by either the registration performance or the registration speed. In this paper, we propose a novel unsupervised deep learning-based registration framework. The proposed method embeds an affine transformation network in a deformable transformation network, which can not only improves the multi-contrast MR image registration performance but also reduces the time requirement for the registration process. In addition, a dual consistency strategy is proposed to achieve bi-directional image registrations so that the robustness of the method can be enhanced. To optimize the framework, we also developed a joint loss function combining the mutual information loss with an elaborately designed prior knowledge-based background suppressing loss. Compared to state-of-the-art registration methods, our framework achieves the best performance with a Dice score of 0.8397. Our method is also 10 times faster than the most competitive method (SyN) when testing on a CPU. In addition, our method can maintain the performance when handling different tasks, while comparison methods show large performance degradations.

Our developed method is not limited to multi-contrast MR image registrations. It can also be applied to unimodal or other multi-modal image registration tasks with modifications. Furthermore, accurate and efficient registration algorithms can be employed in the development of learning-based methods when human annotations are expensive to obtain and reduced reliance on annotations is necessary. For example, the proposed method can be easily extended to ATLAS-based segmentation tasks. In the future, we expect to further develop the proposed method to accommodate multi-modal image registrations such as those from CT to MR images. Overall, our method presents encouraging potentials in assisting intelligent medical data analysis.

\ifCLASSOPTIONcaptionsoff
  \newpage
\fi


%
\bibliographystyle{IEEEtran}
\bibliography{egbib} 

\begin{thebibliography}{10}
\providecommand{\url}[1]{#1}
\csname url@samestyle\endcsname
\providecommand{\newblock}{\relax}
\providecommand{\bibinfo}[2]{#2}
\providecommand{\BIBentrySTDinterwordspacing}{\spaceskip=0pt\relax}
\providecommand{\BIBentryALTinterwordstretchfactor}{4}
\providecommand{\BIBentryALTinterwordspacing}{\spaceskip=\fontdimen2\font plus
\BIBentryALTinterwordstretchfactor\fontdimen3\font minus
  \fontdimen4\font\relax}
\providecommand{\BIBforeignlanguage}[2]{{%
\expandafter\ifx\csname l@#1\endcsname\relax
\typeout{** WARNING: IEEEtran.bst: No hyphenation pattern has been}%
\typeout{** loaded for the language `#1'. Using the pattern for}%
\typeout{** the default language instead.}%
\else
\language=\csname l@#1\endcsname
\fi
#2}}
\providecommand{\BIBdecl}{\relax}
\BIBdecl

\bibitem{krebs2019learning}
J.~Krebs, H.~Delingette, B.~Mailh{\'e}, N.~Ayache, and T.~Mansi, ``Learning a
  probabilistic model for diffeomorphic registration,'' \emph{IEEE transactions
  on medical imaging}, vol.~38, no.~9, pp. 2165--2176, 2019.

\bibitem{balakrishnan2018unsupervised}
G.~Balakrishnan, A.~Zhao, M.~R. Sabuncu, J.~Guttag, and A.~V. Dalca, ``An
  unsupervised learning model for deformable medical image registration,'' in
  \emph{Proceedings of the IEEE conference on computer vision and pattern
  recognition}, 2018, pp. 9252--9260.

\bibitem{dalca2018unsupervised}
A.~V. Dalca, G.~Balakrishnan, J.~Guttag, and M.~R. Sabuncu, ``Unsupervised
  learning for fast probabilistic diffeomorphic registration,'' in
  \emph{International Conference on Medical Image Computing and
  Computer-Assisted Intervention}.\hskip 1em plus 0.5em minus 0.4em\relax
  Springer, 2018, pp. 729--738.

\bibitem{balakrishnan2019voxelmorph}
G.~Balakrishnan, A.~Zhao, M.~R. Sabuncu, J.~Guttag, and A.~V. Dalca,
  ``Voxelmorph: a learning framework for deformable medical image
  registration,'' \emph{IEEE transactions on medical imaging}, vol.~38, no.~8,
  pp. 1788--1800, 2019.

\bibitem{dalca2019unsupervised}
A.~V. Dalca, G.~Balakrishnan, J.~Guttag, and M.~R. Sabuncu, ``Unsupervised
  learning of probabilistic diffeomorphic registration for images and
  surfaces,'' \emph{Medical image analysis}, vol.~57, pp. 226--236, 2019.

\bibitem{Wang2017Learning}
S.~Wang, S.~Tan, Y.~Gao, Q.~Liu, and D.~Liang, ``Learning joint-sparse codes
  for calibration-free parallel mr imaging (lindberg),'' \emph{IEEE
  Transactions on Medical Imaging}, vol.~PP, no.~99, pp. 1--1, 2017.

\bibitem{chaisaowong2018automated}
K.~Chaisaowong and M.~Jiang, ``An automated 3d-atlas-based registration towards
  the anatomical segmentation of pulmonary pleural surface,'' in \emph{2018
  International ECTI Northern Section Conference on Electrical, Electronics,
  Computer and Telecommunications Engineering (ECTI-NCON)}.\hskip 1em plus
  0.5em minus 0.4em\relax IEEE, 2018, pp. 85--88.

\bibitem{marstal2019continuous}
K.~Marstal, F.~Berendsen, N.~Dekker, M.~Staring, and S.~Klein, ``The continuous
  registration challenge: Evaluation-as-a-service for medical image
  registration algorithms,'' in \emph{2019 IEEE 16th International Symposium on
  Biomedical Imaging (ISBI 2019)}.\hskip 1em plus 0.5em minus 0.4em\relax IEEE,
  2019, pp. 1399--1402.

\bibitem{cao2018region}
X.~Cao, J.~Yang, Y.~Gao, Q.~Wang, and D.~Shen, ``Region-adaptive deformable
  registration of {CT/MRI} pelvic images via learning-based image synthesis,''
  \emph{IEEE Transactions on Image Processing}, vol.~27, no.~7, pp. 3500--3512,
  2018.

\bibitem{cao2018deformable}
X.~Cao, J.~Yang, J.~Zhang, Q.~Wang, P.-T. Yap, and D.~Shen, ``Deformable image
  registration using a cue-aware deep regression network,'' \emph{IEEE
  Transactions on Biomedical Engineering}, vol.~65, no.~9, pp. 1900--1911,
  2018.

\bibitem{zhao2017novel}
Y.~Zhao, S.~Zhang, H.~Chen, W.~Zhang, J.~Lv, X.~Jiang, D.~Shen, and T.~Liu, ``A
  novel framework for groupwise registration of {fMRI} images based on common
  functional networks,'' in \emph{2017 IEEE 14th International Symposium on
  Biomedical Imaging (ISBI 2017)}.\hskip 1em plus 0.5em minus 0.4em\relax IEEE,
  2017, pp. 485--489.

\bibitem{konig2015parallel}
L.~K{\"o}nig, A.~Derksen, M.~Hallmann, and N.~Papenberg, ``Parallel and memory
  efficient multimodal image registration for radiotherapy using normalized
  gradient fields,'' in \emph{2015 IEEE 12th international symposium on
  biomedical imaging (ISBI)}.\hskip 1em plus 0.5em minus 0.4em\relax IEEE,
  2015, pp. 734--738.

\bibitem{li2017pixel}
S.~Li, X.~Kang, L.~Fang, J.~Hu, and H.~Yin, ``Pixel-level image fusion: A
  survey of the state of the art,'' \emph{information Fusion}, vol.~33, pp.
  100--112, 2017.

\bibitem{stejskal1965spin}
E.~O. Stejskal and J.~E. Tanner, ``Spin diffusion measurements: spin echoes in
  the presence of a time-dependent field gradient,'' \emph{The journal of
  chemical physics}, vol.~42, no.~1, pp. 288--292, 1965.

\bibitem{basser1994mr}
P.~J. Basser, J.~Mattiello, and D.~LeBihan, ``Mr diffusion tensor spectroscopy
  and imaging,'' \emph{Biophysical journal}, vol.~66, no.~1, pp. 259--267,
  1994.

\bibitem{lauterbur1973image}
P.~C. Lauterbur, ``Image formation by induced local interactions: examples
  employing nuclear magnetic resonance,'' \emph{nature}, vol. 242, no. 5394,
  pp. 190--191, 1973.

\bibitem{hill2001medical}
D.~L. Hill, P.~G. Batchelor, M.~Holden, and D.~J. Hawkes, ``Medical image
  registration,'' \emph{Physics in medicine \& biology}, vol.~46, no.~3, p.~R1,
  2001.

\bibitem{brown1992survey}
L.~G. Brown, ``A survey of image registration techniques,'' \emph{ACM computing
  surveys (CSUR)}, vol.~24, no.~4, pp. 325--376, 1992.

\bibitem{van1999automated}
K.~Van~Leemput, F.~Maes, D.~Vandermeulen, and P.~Suetens, ``Automated
  model-based bias field correction of mr images of the brain,'' \emph{IEEE
  transactions on medical imaging}, vol.~18, no.~10, pp. 885--896, 1999.

\bibitem{dawant2002non}
B.~M. Dawant, ``Non-rigid registration of medical images: purpose and methods,
  a short survey,'' in \emph{Proceedings IEEE International Symposium on
  Biomedical Imaging}.\hskip 1em plus 0.5em minus 0.4em\relax IEEE, 2002, pp.
  465--468.

\bibitem{reese2003reduction}
T.~G. Reese, O.~Heid, R.~Weisskoff, and V.~Wedeen, ``Reduction of
  eddy-current-induced distortion in diffusion mri using a twice-refocused spin
  echo,'' \emph{Magnetic Resonance in Medicine: An Official Journal of the
  International Society for Magnetic Resonance in Medicine}, vol.~49, no.~1,
  pp. 177--182, 2003.

\bibitem{bajcsy1989multiresolution}
R.~Bajcsy and S.~Kova{\v{c}}i{\v{c}}, ``Multiresolution elastic matching,''
  \emph{Computer vision, graphics, and image processing}, vol.~46, no.~1, pp.
  1--21, 1989.

\bibitem{shen2002hammer}
D.~Shen and C.~Davatzikos, ``Hammer: hierarchical attribute matching mechanism
  for elastic registration,'' \emph{IEEE transactions on medical imaging},
  vol.~21, no.~11, pp. 1421--1439, 2002.

\bibitem{beg2005computing}
M.~F. Beg, M.~I. Miller, A.~Trouv{\'e}, and L.~Younes, ``Computing large
  deformation metric mappings via geodesic flows of diffeomorphisms,''
  \emph{International journal of computer vision}, vol.~61, no.~2, pp.
  139--157, 2005.

\bibitem{hart2009optimal}
G.~L. Hart, C.~Zach, and M.~Niethammer, ``An optimal control approach for
  deformable registration,'' in \emph{2009 IEEE Computer Society Conference on
  Computer Vision and Pattern Recognition Workshops}.\hskip 1em plus 0.5em
  minus 0.4em\relax IEEE, 2009, pp. 9--16.

\bibitem{vercauteren2009diffeomorphic}
T.~Vercauteren, X.~Pennec, A.~Perchant, and N.~Ayache, ``Diffeomorphic demons:
  Efficient non-parametric image registration,'' \emph{NeuroImage}, vol.~45,
  no.~1, pp. S61--S72, 2009.

\bibitem{chen2013large}
Z.~Chen, H.~Jin, Z.~Lin, S.~Cohen, and Y.~Wu, ``Large displacement optical flow
  from nearest neighbor fields,'' in \emph{Proceedings of the IEEE Conference
  on Computer Vision and Pattern Recognition}, 2013, pp. 2443--2450.

\bibitem{wulff2015efficient}
J.~Wulff and M.~J. Black, ``Efficient sparse-to-dense optical flow estimation
  using a learned basis and layers,'' in \emph{Proceedings of the IEEE
  Conference on Computer Vision and Pattern Recognition}, 2015, pp. 120--130.

\bibitem{rueckert1999nonrigid}
D.~Rueckert, L.~I. Sonoda, C.~Hayes, D.~L. Hill, M.~O. Leach, and D.~J. Hawkes,
  ``Nonrigid registration using free-form deformations: application to breast
  mr images,'' \emph{IEEE transactions on medical imaging}, vol.~18, no.~8, pp.
  712--721, 1999.

\bibitem{thirion1998image}
J.-P. Thirion, ``Image matching as a diffusion process: an analogy with
  maxwell's demons,'' \emph{Medical image analysis}, vol.~2, no.~3, pp.
  243--260, 1998.

\bibitem{wang2005validation}
H.~Wang, L.~Dong, J.~O'Daniel, R.~Mohan, A.~S. Garden, K.~K. Ang, D.~A. Kuban,
  M.~Bonnen, J.~Y. Chang, and R.~Cheung, ``Validation of an accelerated
  ‘demons’ algorithm for deformable image registration in radiation
  therapy,'' \emph{Physics in Medicine \& Biology}, vol.~50, no.~12, p. 2887,
  2005.

\bibitem{vercauteren2007non}
T.~Vercauteren, X.~Pennec, A.~Perchant, and N.~Ayache, ``Non-parametric
  diffeomorphic image registration with the demons algorithm,'' in
  \emph{International Conference on Medical Image Computing and
  Computer-Assisted Intervention}.\hskip 1em plus 0.5em minus 0.4em\relax
  Springer, 2007, pp. 319--326.

\bibitem{shen2019region}
Z.~Shen, F.-X. Vialard, and M.~Niethammer, ``Region-specific diffeomorphic
  metric mapping,'' in \emph{Advances in Neural Information Processing
  Systems}, 2019, pp. 1098--1108.

\bibitem{shen2019networks}
Z.~Shen, X.~Han, Z.~Xu, and M.~Niethammer, ``Networks for joint affine and
  non-parametric image registration,'' in \emph{Proceedings of the IEEE
  Conference on Computer Vision and Pattern Recognition}, 2019, pp. 4224--4233.

\bibitem{avants2008symmetric}
B.~B. Avants, C.~L. Epstein, M.~Grossman, and J.~C. Gee, ``Symmetric
  diffeomorphic image registration with cross-correlation: evaluating automated
  labeling of elderly and neurodegenerative brain,'' \emph{Medical image
  analysis}, vol.~12, no.~1, pp. 26--41, 2008.

\bibitem{krebs2018learning}
J.~Krebs, T.~Mansi, B.~Mailh{\'e}, N.~Ayache, and H.~Delingette, ``Learning
  structured deformations using diffeomorphic registration,'' \emph{arXiv
  preprint arXiv:1804.07172}, 2018.

\bibitem{krebs2017robust}
J.~Krebs, T.~Mansi, H.~Delingette, L.~Zhang, F.~C. Ghesu, S.~Miao, A.~K. Maier,
  N.~Ayache, R.~Liao, and A.~Kamen, ``Robust non-rigid registration through
  agent-based action learning,'' in \emph{International Conference on Medical
  Image Computing and Computer-Assisted Intervention}.\hskip 1em plus 0.5em
  minus 0.4em\relax Springer, 2017, pp. 344--352.

\bibitem{liao2017artificial}
R.~Liao, S.~Miao, P.~de~Tournemire, S.~Grbic, A.~Kamen, T.~Mansi, and
  D.~Comaniciu, ``An artificial agent for robust image registration,'' in
  \emph{Thirty-First AAAI Conference on Artificial Intelligence}, 2017.

\bibitem{ma2017multimodal}
K.~Ma, J.~Wang, V.~Singh, B.~Tamersoy, Y.-J. Chang, A.~Wimmer, and T.~Chen,
  ``Multimodal image registration with deep context reinforcement learning,''
  in \emph{International Conference on Medical Image Computing and
  Computer-Assisted Intervention}.\hskip 1em plus 0.5em minus 0.4em\relax
  Springer, 2017, pp. 240--248.

\bibitem{miao2018dilated}
S.~Miao, S.~Piat, P.~Fischer, A.~Tuysuzoglu, P.~Mewes, T.~Mansi, and R.~Liao,
  ``Dilated fcn for multi-agent 2d/3d medical image registration,'' in
  \emph{Thirty-Second AAAI Conference on Artificial Intelligence}, 2018.

\bibitem{de2019deep}
B.~D. de~Vos, F.~F. Berendsen, M.~A. Viergever, H.~Sokooti, M.~Staring, and
  I.~I{\v{s}}gum, ``A deep learning framework for unsupervised affine and
  deformable image registration,'' \emph{Medical image analysis}, vol.~52, pp.
  128--143, 2019.

\bibitem{li2018non}
H.~Li and Y.~Fan, ``Non-rigid image registration using self-supervised fully
  convolutional networks without training data,'' in \emph{2018 IEEE 15th
  International Symposium on Biomedical Imaging (ISBI 2018)}.\hskip 1em plus
  0.5em minus 0.4em\relax IEEE, 2018, pp. 1075--1078.

\bibitem{haskins2020deep}
G.~Haskins, U.~Kruger, and P.~Yan, ``Deep learning in medical image
  registration: a survey,'' \emph{Machine Vision and Applications}, vol.~31,
  no.~1, p.~8, 2020.

\bibitem{jaderberg2015spatial}
\BIBentryALTinterwordspacing
M.~Jaderberg, K.~Simonyan, A.~Zisserman, and k.~kavukcuoglu, ``Spatial
  transformer networks,'' in \emph{Advances in Neural Information Processing
  Systems 28}, C.~Cortes, N.~D. Lawrence, D.~D. Lee, M.~Sugiyama, and
  R.~Garnett, Eds.\hskip 1em plus 0.5em minus 0.4em\relax Curran Associates,
  Inc., 2015, pp. 2017--2025. [Online]. Available:
  \url{http://papers.nips.cc/paper/5854-spatial-transformer-networks.pdf}
\BIBentrySTDinterwordspacing

\bibitem{yoo2017ssemnet}
I.~Yoo, D.~G. Hildebrand, W.~F. Tobin, W.-C.~A. Lee, and W.-K. Jeong,
  ``ssemnet: Serial-section electron microscopy image registration using a
  spatial transformer network with learned features,'' in \emph{Deep Learning
  in Medical Image Analysis and Multimodal Learning for Clinical Decision
  Support}.\hskip 1em plus 0.5em minus 0.4em\relax Springer, 2017, pp.
  249--257.

\bibitem{krebs2018unsupervised}
J.~Krebs, T.~Mansi, B.~Mailh{\'e}, N.~Ayache, and H.~Delingette, ``Unsupervised
  probabilistic deformation modeling for robust diffeomorphic registration,''
  in \emph{Deep Learning in Medical Image Analysis and Multimodal Learning for
  Clinical Decision Support}.\hskip 1em plus 0.5em minus 0.4em\relax Springer,
  2018, pp. 101--109.

\bibitem{maes1997multimodality}
F.~Maes, A.~Collignon, D.~Vandermeulen, G.~Marchal, and P.~Suetens,
  ``Multimodality image registration by maximization of mutual information,''
  \emph{IEEE transactions on Medical Imaging}, vol.~16, no.~2, pp. 187--198,
  1997.

\bibitem{li2018multi}
Z.~Li, F.~Huang, J.~Zhang, B.~Dashtbozorg, S.~Abbasi-Sureshjani, Y.~Sun,
  X.~Long, Q.~Yu, B.~ter Haar~Romeny, and T.~Tan, ``Multi-modal and
  multi-vendor retina image registration,'' \emph{Biomedical optics express},
  vol.~9, no.~2, pp. 410--422, 2018.

\bibitem{ceranka2018registration}
J.~Ceranka, M.~Polfliet, F.~Lecouvet, N.~Michoux, J.~de~Mey, and
  J.~Vandemeulebroucke, ``Registration strategies for multi-modal whole-body
  {MRI} mosaicing,'' \emph{Magnetic resonance in medicine}, vol.~79, no.~3, pp.
  1684--1695, 2018.

\bibitem{cao2017dual}
X.~Cao, J.~Yang, Y.~Gao, Y.~Guo, G.~Wu, and D.~Shen, ``Dual-core steered
  non-rigid registration for multi-modal images via bi-directional image
  synthesis,'' \emph{Medical image analysis}, vol.~41, pp. 18--31, 2017.

\bibitem{cao2018deep}
X.~Cao, J.~Yang, L.~Wang, Z.~Xue, Q.~Wang, and D.~Shen, ``Deep learning based
  inter-modality image registration supervised by intra-modality similarity,''
  in \emph{International Workshop on Machine Learning in Medical
  Imaging}.\hskip 1em plus 0.5em minus 0.4em\relax Springer, 2018, pp. 55--63.

\bibitem{dong2002affine}
P.~Dong and N.~P. Galatsanos, ``Affine transformation resistant watermarking
  based on image normalization,'' in \emph{Proceedings. International
  Conference on Image Processing}, vol.~3.\hskip 1em plus 0.5em minus
  0.4em\relax IEEE, 2002, pp. 489--492.

\bibitem{wang2020lt}
S.~Wang, S.~Cao, D.~Wei, R.~Wang, K.~Ma, L.~Wang, D.~Meng, and Y.~Zheng,
  ``Lt-net: Label transfer by learning reversible voxel-wise correspondence for
  one-shot medical image segmentation,'' in \emph{Proceedings of the IEEE/CVF
  Conference on Computer Vision and Pattern Recognition}, 2020, pp. 9162--9171.

\bibitem{vajda1989theory}
I.~Vajda, \emph{Theory of statistical inference and information}.\hskip 1em
  plus 0.5em minus 0.4em\relax Kluwer Academic Pub, 1989, vol.~11.

\bibitem{ronneberger2015u}
O.~Ronneberger, P.~Fischer, and T.~Brox, ``U-net: Convolutional networks for
  biomedical image segmentation,'' in \emph{International Conference on Medical
  image computing and computer-assisted intervention}.\hskip 1em plus 0.5em
  minus 0.4em\relax Springer, 2015, pp. 234--241.

\bibitem{avants2011reproducible}
B.~B. Avants, N.~J. Tustison, G.~Song, P.~A. Cook, A.~Klein, and J.~C. Gee, ``A
  reproducible evaluation of {ANTs} similarity metric performance in brain
  image registration,'' \emph{Neuroimage}, vol.~54, no.~3, pp. 2033--2044,
  2011.

\end{thebibliography}

\end{document}